\documentclass[12pt]{article}

\usepackage[OT1]{fontenc} 
\usepackage{graphicx}
\usepackage{amssymb,amsmath}
\usepackage{color, colortbl}
\usepackage{mathrsfs}
\usepackage[flushleft]{threeparttable}
\usepackage[normalem]{ulem}
\usepackage{afterpage}
\usepackage{caption}
\makeatletter
\@fpsep\textheight
\makeatother

\newcommand{\araa}{Ann. Rev. A\&A}
\newcommand{\apjl}{ApJL}
\newcommand{\aap}{A\&A}
\newcommand{\aaps}{Astron. Astrophys. Suppl. Ser.}
\newcommand{\nat} {Nature}
\newcommand{\mnras} {MNRAS.}
\newcommand{\apj}{ApJ}
\newcommand{\pasp}{PASP.}
 
\newcommand{\apjs}{ApJS}

\newcommand{\aapr}{AApR}

\newcommand{\pasj}{PASJ}
\newcommand{\pasa}{PASA}

\newenvironment{myAbstract}{%
\begin{quote} \bf}
{\end{quote}}

\newcounter{lastnote}

\title{Observational identification of a sample of likely recent Common-Envelope Events} 

\author
{Theo Khouri$^{1\ast\dagger}$, Wouter H. T. Vlemmings$^{1\dagger}$,
Daniel Tafoya$^{1\dagger}$,\\
Andr\'es F. P\'erez-S\'anchez$^{2}$,
Carmen S\'anchez Contreras$^{3}$, \\
Jos\'e F. G\'omez$^4$, Hiroshi Imai$^{5,6,7}$, Raghvendra Sahai$^8$ \\
\\
\normalsize{$^{1}$Department of Space, Earth and Environment, Chalmers University of Technology,}\\
\normalsize{Onsala Space Observatory, 439~92 Onsala, Sweden}\\

\normalsize{$^{2}$Leiden Observatory, Leiden University,}\\
\normalsize{Niels Bohrweg 2, 2333 CA Leiden, The Netherlands}\\

\normalsize{$^3$ Centro de Astrobiolog\'ia (CSIC-INTA), Camino Bajo del Castillo s/n,}\\
\normalsize{Urb. Villafranca del Castillo, Villanueva de la Ca\~{n}ada, E-28691 Madrid, Spain}\\

\normalsize{$^4$ Instituto de Astrof\'isica de Andaluc\'ia, CSIC,}\\
\normalsize{Glorieta de la Astronom\'ia s/n, E-18008 Granada, Spain}\\

\normalsize{$^5$ Amanogawa Galaxy Astronomy Research Center, Graduate School of Science and Engineering,}\\
\normalsize{Kagoshima University , 1-21-35 Korimoto, Kagoshima, Kagoshima 890-0065, Japan}\\

\normalsize{$^6$ Center for General Education, Institute for Comprehensive Education,}\\
\normalsize{Kagoshima University, 1-21-30 Korimoto, Kagoshima, Kagoshima 890-0065, Japan}\\

\normalsize{$^7$ Department of Physics and Astronomy, Graduate School of Science and Engineering,}\\
\normalsize{Kagoshima University, 1-21-35 Korimoto, Kagoshima, Kagoshima 890-0065, Japan}\\

\normalsize{$^8$ Jet Propulsion Laboratory, California Institute of Technology,}\\
\normalsize{4800 Oak Grove Drive, Pasadena, CA 91109, USA}\\

\\
\normalsize{$^\ast$ Corresponding author. E-mail:  theo.khouri@chalmers.se.}\\

\normalsize{$^\dagger$ These authors contributed equally to this work}
}

\date{\today}

\begin{document} 


\baselineskip24pt


\maketitle 
\pagebreak

\begin{myAbstract}
One of  the  most  poorly  understood  stellar  evolutionary  paths  is  that  of  binary systems undergoing common-envelope evolution, when the envelope of a  giant star  engulfs  the  orbit of a companion. Although this interaction leads to a great variety of astrophysical systems, direct empirical studies are difficult because few objects experiencing common-envelope evolution are known.  We present ALMA observations towards sources known as water fountains that reveal they had low initial masses ($<4~{\rm M}_\odot$) and ejected a significant fraction of it over less than a few hundred years. The only mechanism able to explain such rapid mass ejection is common-envelope evolution. Our calculations show that the water-fountain sample accounts for a large fraction of the systems in our Galaxy which have just experienced the common-envelope phase. Since water-fountain sources show characteristic fast bipolar outflows, outflows and jets likely play an important role right before, during or immediately after the common-envelope phase.
\end{myAbstract}

Binary stellar systems where both companions temporarily orbit within
a shared envelope undergo common envelope evolution (CEE). The common
envelope (CE) phase is thought to be a crucial stage in the evolution
of a large number of binary stars \cite{Paczynski1976,Iben1993,Ivanova2013}. Specifically, it is one of the most likely
formation pathways of Type Ia supernova progenitors \cite{Iben1984}. The CEE is also strongly linked to the formation of many
planetary nebulae (PNe) with a wide variety of shapes \cite{Han1995} as well as the
formation of double neutron star systems \cite{Taam2000}. Most gravitational wave sources will have undergone CEE
\cite{Belczynski2002}. An evolutionary path analogous to CEE might be important even for single star systems, because the interaction with massive
planets could also be considered a form of CEE \cite{Villaver2009}.

The short CE phase often occurs towards the final stages of stellar
evolution, when the binary orbit becomes unstable due to, for example,
mass transfer that is initiated when one of the stars fills its Roche
Lobe. When a companion enters the envelope of the Roche-lobe filling
star, a CE is formed that will cause drag on the stars orbiting
within. As a result, the orbit decays further and orbital energy and
angular momentum is transferred to the CE. The end point of this short
evolutionary stage is either a stellar merger, or a compact binary if
the envelope is ejected prior to merging.

Despite the unquestioned importance of CEE, this is
one of the most poorly understood phases of stellar evolution. Current
models still have problems ejecting the CE \cite{Ivanova2013}
and require the inclusion of additional physics such as hydrogen/helium recombination
\cite{Ivanova2015}, mass accretion and associated jets \cite{Soker2014,MorenoMendez2017,Chamandy2018}, and/or dust
formation \cite{Glanz2018}. The uncertainties are
exacerbated by the difficulty in observing the CE phase because of its
short time scale, from days to years depending on the stage of CE
interaction. Although the number of likely post-CE systems that are
observed is steadily increasing \cite{Sahai2017b,Olofsson2019,Kamisnki2021}, most sources associated with CEE left the CE phase a
long time ago \cite{Jones2020}. With the likely exception of the
class of luminous red novae with characteristics consistent with
CE events \cite{Ivanova2013Sci, Howitt2020}, recent CE systems remain elusive, and direct measurements of the mass and velocity of the ejecta are rare. As
a result, CE simulations can only be compared with highly evolved
descendants, which means that many physical characteristics related to
for example the action of jets and the presence of dust, as well as
the ejection efficiency remains poorly constrained.

\subsection*{Results}
We have identified a class of objects which are most likely
experiencing CEE or have experienced it in the last $\lesssim 200$~years. These systems are known
as water fountains (WFs) and are enshrouded in optically thick
dusty envelopes currently being excavated by nascent jets. The characteristic high-velocity (often at $\gtrsim 100$~km/s) water maser emission that
arises from the interaction between the jet and the circumstellar envelope makes water-fountain (WF) systems relatively easy to detect. In at least three of the WFs, the jets are shown to be related to the presence of magnetic fields \cite{Vlemmings2006,PerezSanchez2013,Suarez2015}.
High-angular-resolution observations reveal
the bulk of the gas to be in slowly expanding ($\sim 20$~km/s) envelopes, often tori-like,
distributed mostly in a perpendicular direction to the bipolar lobes sculpted by the jets
\cite{Sahai2017,Gomez2018,Tafoya2020}.

The nature of WFs has remained controversial since their discovery.
Until now, they were inferred to be relatively massive stars ($M \gtrsim 5$ M$_\odot$) at the tip of the asymptotic giant branch (AGB) or already in the post-AGB phase. This is because of the high mass-loss rates inferred ($\sim 10^{-4}$~M$_\odot$~year$^{-1}$ \cite{Rizzo2013}) and the low values observed for 
the $^{12}$CO/$^{13}$CO line ratios \cite{Imai2012}.
However, this interpretation is inadequate in several ways. First, since we observe the WF phase during a short time ($\lesssim200$~yr), the number of WFs known (15 sources) makes 
it statistically very unlikely that these stars are more massive than a few solar masses (see Methods). Moreover, under this paradigm the circumstellar envelopes are expected to be very extended and mostly spherical.
However, studies of the spectral energy distributions of these sources show that the observations cannot be reproduced using spherical 
symmetric models \cite{Yung2017},
%    This was at first interpreted as a sign that the age of the jets is 
%underestimated, but ages derived from spatially resolved observations of the dense circumstellar 
%cocoon reveal comparable ages for the mass ejections (\textcolor{red}{ am I right? references}).
and observations at high angular resolution of individual systems
show the dust emission to be much more compact than expected for extended envelopes \cite{Sahai2017,Gomez2018,Tafoya2020}.

\begin{figure*}
 	\centering
 	\includegraphics[angle=-0,width=0.6\textwidth]{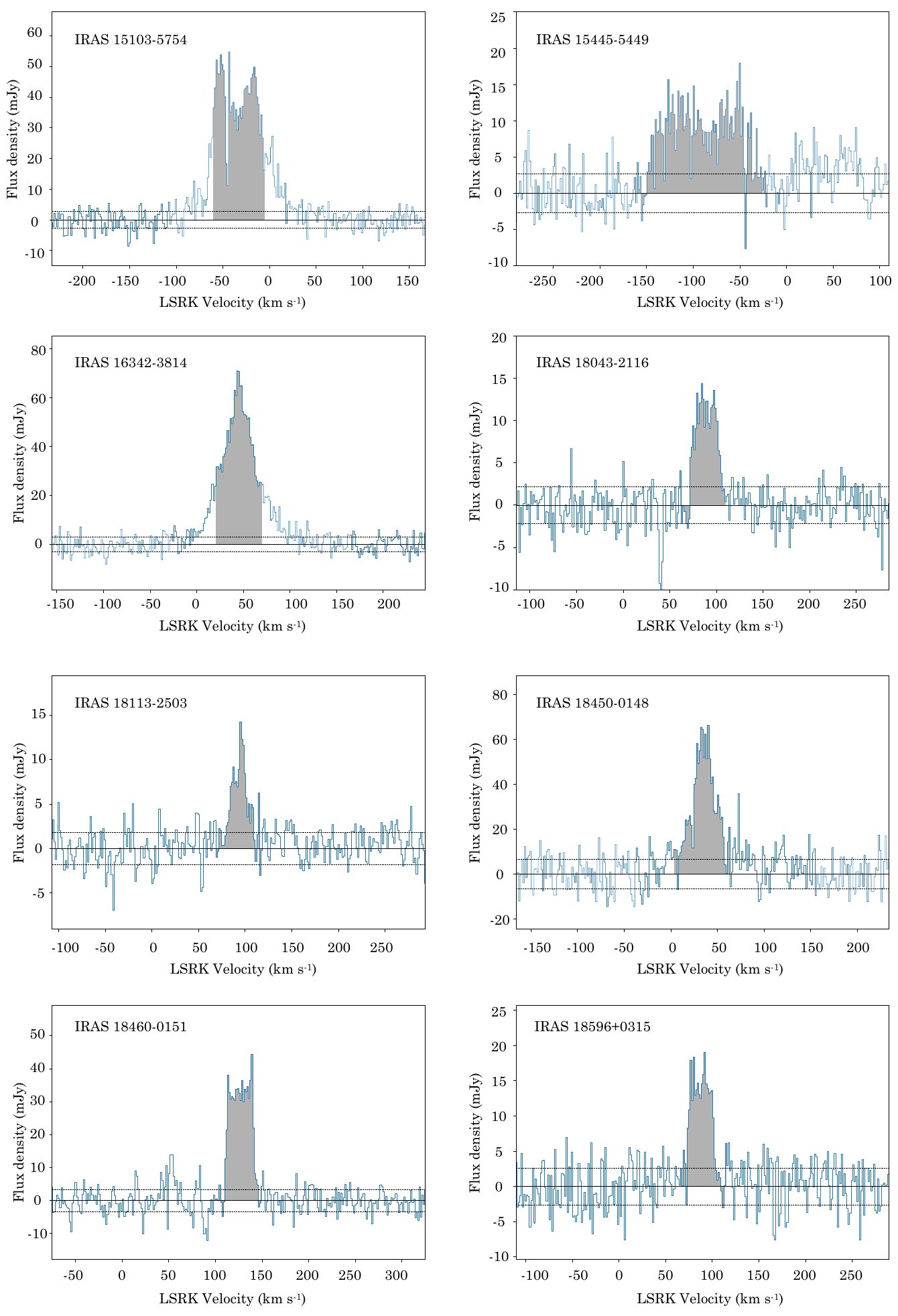}
	\caption{$^{12}$C$^{18}$O spectra towards the sources with detection. The three horizontal lines mark the levels of flux density equal to zero and to the positive and negative values of the root-mean square of the spectra.The shaded area shows the line flux considered in our calculations of the gas mass in the slow outflow (see Methods).}
	\label{fig:C18O}
\end{figure*}

To investigate the nature of WFs, we have used the Atacama Large Millimeter/submillimeter Array (ALMA) to observe the $J=2-1$ transitions of $^{12}$C$^{16}$O, $^{13}$C$^{16}$O, 
$^{12}$C$^{18}$O and the $J=1-0$ transitions of $^{12}$C$^{16}$O 
and $^{12}$C$^{17}$O towards all 15 known WF sources.
{ We detected the $^{12}$C$^{18}$O,~$J=2-1$ line towards eight objects (Table~\ref{Tab:1} and Fig.~\ref{fig:C18O}). The $^{12}$C$^{17}$O,~$J=1-0$ line was detected for all these eight sources but IRAS~15445-5449. This is the first time that emission from the rarer isotopologues of CO are observed towards these WFs (except for IRAS~15103-5754 \cite{Gomez2018}).
As shown in Table~\ref{Tab:IntLines}, $^{12}$C$^{16}$O and $^{13}$C$^{16}$O emission was detected towards all sources with successful observations (the only exception being $^{13}$C$^{16}$O towards IRAS~18139-1816 because of incorrect input coordinates).}
%(Table~\ref{Tab:IntLines}).
%The $^{12}$C$^{16}$O,~$J=2-1$ and $^{13}$C$^{16}$O,~$J=2-1$ lines were detected for all but one source. The only exception is IRAS~18139-1816, which was not observed because the incorrect coordinates caused the source to not be in the field imaged by ALMA.
The $^{12}$C$^{16}$O/$^{13}$C$^{16}$O line ratios have relatively low values (between 2.2 and 6.0), 
which are in agreement with previously reported values \cite{Imai2012,Sahai2017}. For all but three of the sources (IRAS~15103-5457, IRAS~16342-3814, and IRAS~18460-0151), we find the $^{12}$C$^{16}$O emission to be unresolved (Fig.~\ref{fig:12CO_maps}), and there are thus no signs of enhanced mass loss over an extended period of time ($> 1000$~yrs). 
%The only source which shows well-resolved $^{12}$C$^{18}$O emission is IRAS~18460-0151. The emission region translates to mass loss operating over a few thousand years.

\begin{table*}[!ht]
\caption{Physical parameters of the water fountain sources with $^{12}$C$^{18}$O detection.}             % title of Table
\label{Tab:1}      % is used to refer this table in the text
\begin{center}
	\begin{tabular}{ c c c c c c c}
		\hline\hline
		IRAS name & $d$ & $M^{\rm H_2}$ &C$^{16}$O/C$^{17}$O & C$^{17}$O/C$^{18}$O & $\Delta t$ & $\Delta M^{\rm H_2} / \Delta t$ \\
		  & (kpc) & ($M_{\odot}$)& & & (yr) & ($10^{-3}~M_\odot$/yr)  \\
		\hline 
		15103$-$5754 &3.4  & $0.68$ & $278$ & $2.05 \pm 0.33$ & $206$ & $3.3$\\ 
		15445$-$5449 &4.4  & $0.55$ & $>630$ & $<0.93$ & $110$ & $3.8$\\ 
		16342$-$3814 &2.2  & $0.29$ & $1140$ & $0.50 \pm 0.15$ & $150$ & $2.0$\\
		18043$-$2116 &8.2  & $0.62$ & $154$ & $3.7\pm 1.0$ & $195$ & $1.6$ \\ 
		18113$-$2503&12.0  & $0.87$ & $98$ & $5.8 \pm 1.6$ & $65$ & $14$\\ 
		18450$-$0148 &2.2  & $0.24$ & $317$ & $1.8\pm 0.3$ & $95$ & $2.5$\\ 
		18460$-$0151 &3.1  & $0.27$ & $320$ & $1.8\pm 0.4$ & $150$ & $2.0$\\
		18596$+$0315 &8.3  & $0.80$ & $154$ & $3.7 \pm 0.8$ & $<1000$ & $> 0.8$\\ 
		\hline 
	\end{tabular}
\end{center}
\end{table*}

\begin{figure*}[!p]
 	\centering
 	\includegraphics[angle=-0,width=0.99\textwidth]{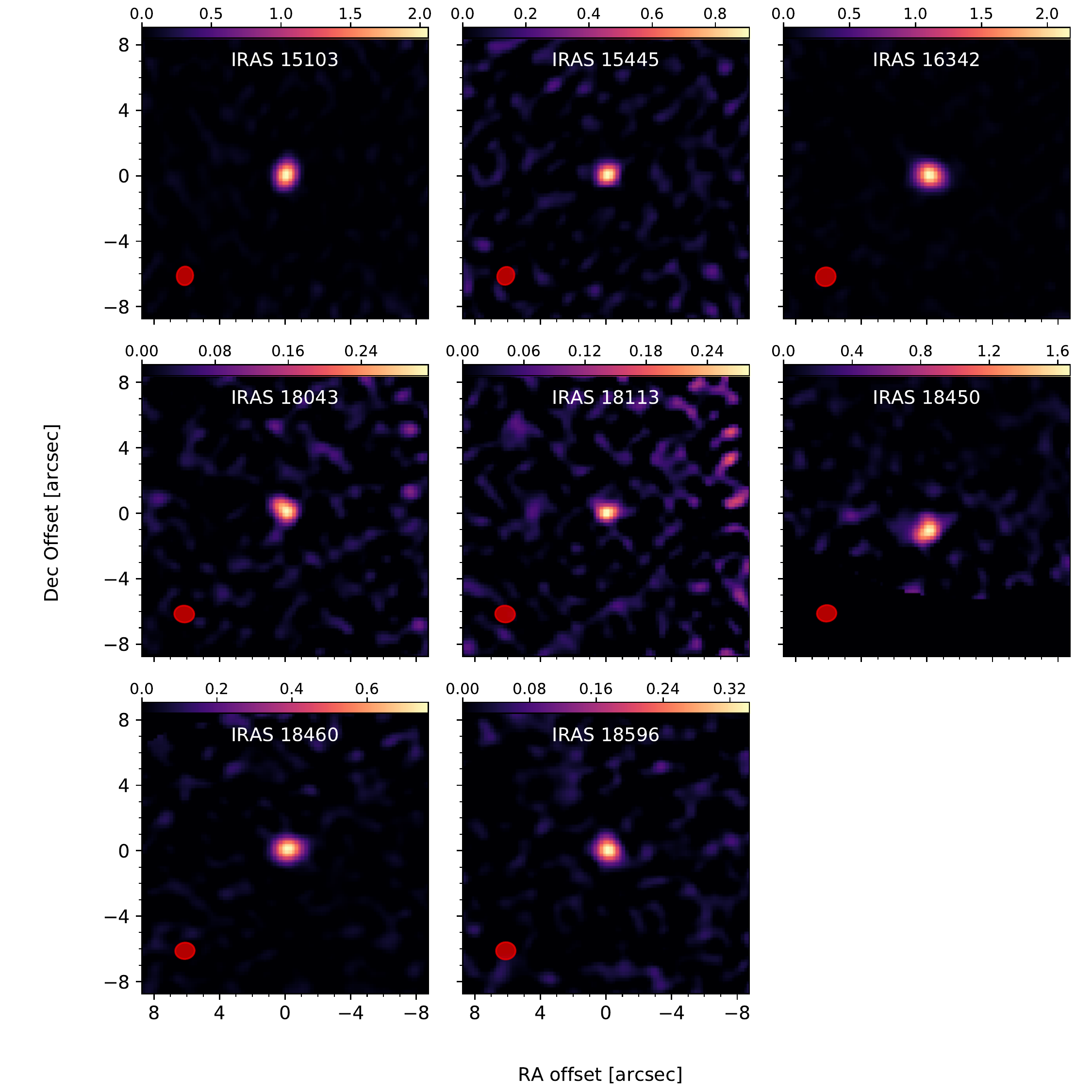}
	\caption{Images of the integrated intensity of the $^{12}$C$^{18}$O,~J=2-1 line towards the water fountain sources with detection, in units of ${\rm Jy\times(km/s)/beam}$.}
	\label{fig:C18O_maps}
\end{figure*}

%These observations shed light on the properties of the central stars and the circumstellar gas, and allow us to investigate the properties of water-fountain sample in a comprehensive and unprecedented manner.
Based on the relative fluxes of the $^{12}$C$^{16}$O and $^{13}$C$^{16}$O lines and on the signal-to-noise ratio
of the obtained spectra, the data are consistent with all 15 sources having comparable 
fluxes of the lines of $^{12}$C$^{17}$O and $^{12}$C$^{18}$O with respect to those of $^{12}$C$^{16}$O and $^{13}$C$^{16}$O. Hence, the non-detections are most likely due to the overall poorer signal-to-noise ratio rather than intrinsically low abundances of the rarer isotopologues.
% Hence, we conclude that the vast majority, and potentially all, WF sources present $^{12}$C$^{18}$O emission.
The implied abundance of $^{18}$O in their circumstellar envelopes is incompatible with these stars 
being intermediate-mass stars ($M \gtrsim 5$~M$_\odot$) at the tip of the AGB { because, in this case, theoretical models predict $^{18}$O abundances orders of magnitude lower than we find \cite{Boothroyd1995,Karakas2016}}. Hence, our observations unequivocally show that WF objects 
must either be less massive, or have had their evolution interrupted. The rarer isotopologue lines are not spatially resolved in our observations (Fig.~\ref{fig:C18O_maps})
but the expansion velocities (Fig.~\ref{fig:C18O})
indicate that these mostly trace gas in the slowly expanding envelope (rather than in the high-velocity
outflows), as was noted for IRAS~15103-5754 \cite{Gomez2018}. The only sources for which we detect the high-velocity outflows in the lines of the rarer isotopologues are IRAS~16342-3814 and IRAS~15103-5754.

The observations allow us to estimate the stellar masses based on the $^{17}$O/$^{18}$O ratio. This ratio is a good tracer of the initial stellar mass because the surface abundance of $^{17}$O is modified
by a mixing process that is strongly dependent on the initial stellar mass. The relevant mixing process takes place much before the AGB,
when these stars evolve through the red giant branch (RGB).
%For stars more massive than about 4~$M_\odot$, the surface abundance of $^{18}$O drops strongly during AGB evolution and the $^{17}$O/$^{18}$O isotopic ratio is expected to reach values order of magnitude larger than we observe.
By assuming optically thin emission and solar composition, we
compare the observed $^{17}$O/$^{18}$O ratios (Table~\ref{Tab:1}) to predictions from theoretical models
\cite{Karakas2014,Abia2017}.
The derived upper limit for IRAS~15445-5449 and $^{17}$O/$^{18}$O ratio for IRAS~16342-3814 imply initial stellar masses $\lesssim 1.4$~M$_\odot$ for these two sources.
%For IRAS~18460−0151 and IRAS~18450−0148, the derived ratios imply initial masses $\sim 1.7$~M$_\odot$, while
For the other six sources with measured $^{17}$O/$^{18}$O ratio, the values 
imply initial masses between 1.8  and 4~M$_\odot$.
Stars in this mass range and solar metallicity are expected to become carbon stars (with carbon abundance larger than that of oxygen) during their 
AGB evolution \cite{Abia2017}, which is markedly different from the oxygen-rich character of WFs. Hence, our observations are not consistent with these six sources having reached the tip of the AGB.
{ Together with the large average mass-loss rates derived below,} this suggests that their evolution was interrupted by a catastrophic mass-loss process. Although the inferred oxygen isotopic ratios and carbon content are consistent with IRAS~15445-5449 and IRAS~16342-3814 being at the tip of the AGB, they are also consistent with them having had an interrupted evolution. This is because no significant changes to the oxygen isotopic ratios or surface carbon abundance are expected for stars in this mass range after the RGB \cite{Karakas2016}. The derivation of the oxygen isotopic ratios and a more extensive discussion on the comparison with theoretical models is given in the Methods.

% \afterpage{\clearpage}

The lower abundance of $^{12}$C$^{18}$O allows us to probe the dense slow-expanding outflow that contains most of the mass, while { we estimate the lines of the main isotopologues to be} very optically thick in that region. Assuming optically thin emission in the $^{12}$C$^{18}$O line, we derive large gas masses of the slowly expanding component of the circumstellar envelopes (Table~\ref{Tab:1}). These are lower limits for the total circumstellar gas masses because the high-velocity outflow is not included in our calculations. Considering
the expansion velocities observed and the typical sizes of the emission region or the inferred lifetimes of the outflows (Fig.~\ref{fig:Andres} and Table~\ref{Tab:isotRatios}), the masses we derive imply very large
average mass-loss rates of $\gtrsim 10^{-3}$~M$_\odot$~year$^{-1}$.
This is larger by at least one order of magnitude than reported in previous studies (with the exception of IRAS~15103-5754 \cite{Gomez2018}). For short single or episodic mass ejections, these average mass-loss rates further underestimate the actual mass-loss rates during the ejection event.

The measured $^{12}$C$^{18}$O line fluxes and the size of the emission regions imply large optical depths in the $^{12}$C$^{16}$O and even $^{13}$C$^{16}$O lines.
This hampers the determination of the $^{12}$C/$^{13}$C isotopic ratios from lines of the main CO isotopologues without detailed radiative 
 transfer modelling. Hence, the low observed values of the $^{12}$C$^{16}$O/$^{13}$C$^{16}$O line ratios in some sources are strongly affected
 by high optical depths rather than necessarily intrinsic low carbon isotopic ratios. 
 %In this way, the carbon isotopic ratios could be consistent with values between 10 and 30, which is expected for stars with $M \lesssim 4$~M$_\odot$ that have not experienced enrichment during the AGB. 
 For IRAS~15445-5449,
  we detected the $J=19-18$ line of Carbonyl Sulfide (OCS) and the O$^{13}$CS isotopologue. The integrated-line-flux ratio is $7.2\pm0.3$, which is a factor of 2.5 larger than the $^{12}$C$^{16}$O/$^{13}$C$^{16}$O line ratio
  observed for this source. { We also find that the flux-density ratio between the $^{12}$C$^{16}$O and the $^{13}$C$^{16}$O lines increases by a factor of five from the line centre to the wings for IRAS~15103-5754, IRAS~15445-5449, and IRAS~16342-3814 (the three sources with lines observed with highest signal-to-noise ratio), as expected for optically thick lines.  This illustrates the strong effect of optical depth on the $^{12}$C$^{16}$O line and on the $^{12}$C$^{16}$O/$^{13}$C$^{16}$O integrated line ratios.}
  %A value of $\sim 7$ for the $^{12}$C/$^{13}$C ratio is still lower than expected for an AGB star with an initial mass $< 4~M_\odot$. However, it is not clear whether the observed line ratio between the OCS isotopologues reflects the intrinsic carbon isotopic ratio, and we conclude that more research is required to understand the carbon isotopic ratios in WFs.

 Considering the oxygen isotopic ratios, the compact emission region, the large circumstellar masses inferred for the sources with observed $^{12}$C$^{18}$O emission (Table~\ref{Tab:1}), and the expansion velocities (as discussed in the Methods), the most plausible scenario is that the majority of
the WF systems are the very recent outcome of CEE, in which a large fraction of the envelopes was ejected. For the objects with detected $^{12}$C$^{18}$O emission, the average mass-loss rates we derive are at least about two orders of magnitude
higher than expected for AGB stars with initial masses $\lesssim 4$~M$_\odot$ \cite{Bloecker1995}.
This is very significant given our estimated uncertainty on the average mass-loss rate of a factor $<5$ (Methods).
In fact, not even the more massive AGB stars $\sim 7$~M$_\odot$ are expected to reach mass-loss rates $> 10^{-3}$~M$_\odot$/yr.

For the seven sources with no $^{12}$C$^{18}$O detection, we use the detection limit and the flux of the $^{13}$C$^{16}$O,~$J=2-1$ line to constrain the hydrogen masses. These are converted to average mass-loss rates using ages available in the literature from the fast outflows, since no spatially resolved images of the slow-outflow components are available (see Methods). We find average mass-loss rates
between $10^{-4}$ and a few times $10^{-3}$~M$_\odot$/yr,
which are also too high considering the inferred initial-mass range for WFs. { Moreover, the linear momentum in the slow component of these outflows only matches that supplied by the radiation field of a star with 5000~L$_\odot$ over tens or hundreds thousand of years, a timescale which is inconsistent with the emission regions and expansion velocities.}

To test whether the number of known WFs are consistent with the CEE scenario, we compare the formation rates of WF systems and CE systems (see Methods for more details).
The duration of the WF phase ($\sim 200$~years, Table~\ref{Tab:isotRatios}) and the 15 known sources imply a formation rate of $>0.075$~WF per year in the Milky Way.
Considering the star formation rate of the Milky Way and the observed binary fraction per mass bin, we estimate that $0.15 \pm 0.05$ binary systems per year will evolve through the
CE phase while one of the stars is in the AGB or at the tip of the red giant branch (RGB). This does not consider possible interactions with giant planets, which might also produce a CE-like interaction. Nonetheless, we estimate that the rate of production of CE systems would increase by $<50\%$ even if giant planets are very efficient in triggering envelope ejection through a type of CEE.
Hence, WFs can account for a sizeable fraction of systems with a giant star undergoing CEE in the Galaxy, or even nearly all, if the sample of currently known WFs is significantly incomplete.

The identification of a new class of objects undergoing CEE has major consequences for the study of this poorly understood phase. Our results imply that WF characteristics
can be now used as telltale signs of the CE phase, the most important of which are the fast outflows traced by water maser emission that can be detected from large distances. In fact, the fast jets and outflows in the WF sample highlight the likely importance of jets immediately before, during or immediately after CEE. The identification of new objects belonging to the WF class and the study of WF objects in general will thus help unveil the details of the complex CE phase and provide much-needed insights to advance our understanding of one of the most enigmatic phases of stellar evolution.

\section*{Acknowledgements}

This paper makes use of the following ALMA data:
ADS/JAO.ALMA 2018.1.00250.S and 2016.1.01032.S.
ALMA is a partnership of ESO (representing its member states), NSF (USA)
and NINS (Japan), together with NRC (Canada) and NSC and ASIAA (Taiwan)
and KASI (Republic of Korea), in cooperation with the Republic of Chile. The
Joint ALMA Observatory is operated by ESO, AUI/NRAO and NAOJ. The authors
acknowledge support from the Nordic ALMA Regional Centre (ARC) node based
at the Onsala Space Observatory. The Nordic ARC node and Swedish observations on APEX are funded through the Swedish
Research Council grant No 2017-00648.
This publication is based on data acquired with the Atacama Pathfinder Experiment (APEX) under programme ID 0104.F-9310(A). APEX is a collaboration between the Max-Planck-Institut fur Radioastronomie, the European Southern Observatory, and the Onsala Space Observatory.

% \noindent
% {\bf Funding:}
\noindent TK: Supported by the Swedish Research Council Starting Grant: 2019-03777.\\
WHTV, TK: Supported by the Swedish Research Council Grant: 2014-05713.\\
CSC: Supported by the Spanish MICINN through grant PID2019-105203GB-C22.\\
JFG: supported by
MCIU-AEI through the “Center of Excellence Severo Ochoa” award to the
Instituto de Astrofísica de Andalucía (SEV-2017-0709), from grants AYA2017-84390-C2-1-R and PID2020-114461GB-I00  of AEI
(10.13039/501100011033), co-funded by FEDER, and from Amanogawa Galaxy Astronomy Research Center (AGARC).\\
HI: Supported by the MEXT KAKENHI program (16H02167) and i-LINK+2019 Programme in IAA/CSIC.
JFG, HI: supported by the Invitation Program for Foreign
Researchers of the Japan Society for Promotion of Science (JSPS, grant S14128).\\
%by the Invitation Program for Foreign Researchers of the Japan Society for Promotion of Science (JSPS grant S14128) and 

\section*{Methods}

\subsection*{Observations and data reduction}

In the context of ALMA project 2018.1.00250.S, the full sample of known water fountain (WF) sources was observed
 between October 2018 and March 2020. Two
different tunings were used, to cover CO,~$J=1-0$ and $J=2-1$ lines in ALMA Bands 3 and 6, respectively.
% and the $^{12}$C$^{16}$O,~$J=2-1$ line in ALMA Band 6.
% CO lines in J=1->0  and J=2->1 in ALMA Bands 3 and 6, respectively
The Band 3 observations included
four spectral windows (spws), each with 3840 channels and a bandwidth
of 1.875 GHz centered on 101.00, 102.85, 113.05 and 114.90~GHz rest
frequency. The third and fourth spws include the $^{12}$C$^{17}$O and $^{12}$C$^{16}$O,~$J=1-0$ lines, respectively. The Band 6 data were taken with six spws. Two 1.875 GHz spws
with 3840 channels were centered on 217.82 and 232.90 GHz, the four
further spws had 1920 channels and a bandwidth of 937.50 MHz, and were
centered on 219.46, 220.40, 230.54, and 231.46 GHz. The first, third, and fourth spws include  the $^{12}$C$^{18}$O, $^{13}$C$^{16}$O,  and $^{12}$C$^{16}$O,~$J=2-1$ lines, respectively.  Sources within 15
degrees were grouped together and the typical observing time,
depending on weather conditions, was approximately 16 minutes on
source in Band 3 and 12 minutes on source in Band 6. This resulted in
a line sensitivity at at phase center with 3~km~s$^{-1}$ resolution
of $\sim1.5$~mJy~beam$^{-1}$ (Band~3) and $\sim1.8$~mJy~beam$^{-1}$ (Band~6). 

\begin{figure*}[!p]
 	\centering
 	\includegraphics[angle=-0,width=0.99\textwidth]{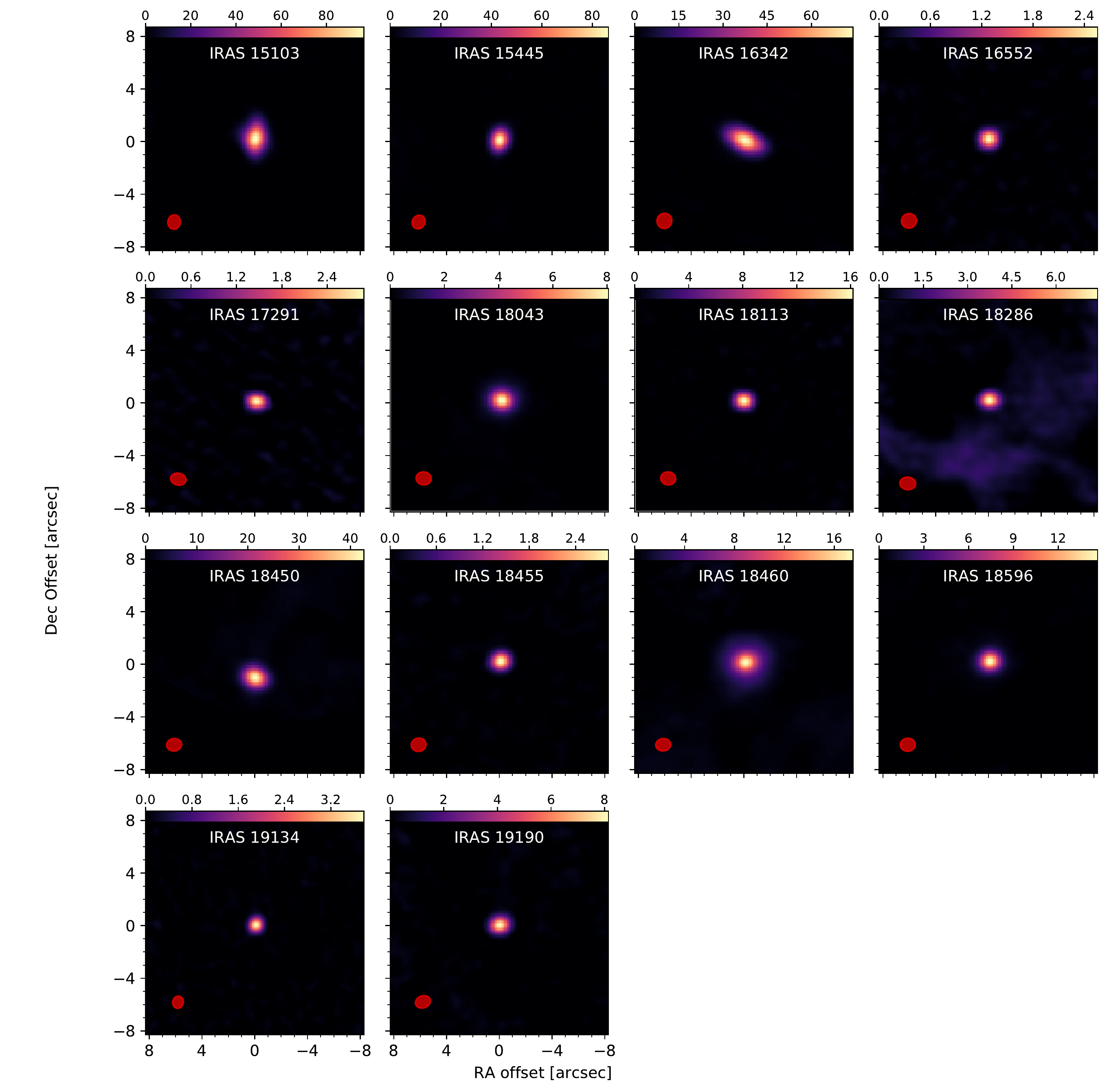}
	\caption{Images of the integrated intensity of the $^{12}$C$^{16}$O,~J=2-1 line towards the water fountain sources, in units of ${\rm Jy\times(km/s)/beam}$}.
	\label{fig:12CO_maps}
\end{figure*}

The data were reduced using ALMA pipeline version 5.4. Subsequently,
a round of phase self-calibration followed by a round of amplitude
self-calibration were performed on the continuum emission of the water fountains (WFs). After
subtracting the continuum emission using the CASA task {\it uvcontsub}, the
synthesized images
were deconvolved with a spectral resolution of $3$~km~s$^{-1}$ in Band
3 and $2.5$~km~s$^{-1}$ in Band 6. Although there were some
differences in the exact antenna configuration, the resulting
resolution was similar for both Bands with a typical beam size of
$\sim1''$ using Briggs weighting (robust parameter equal to 0.5) in the CASA task {\it tclean}. Only
IRAS 19134+2131 was observed at resolutions of $\sim0.6''$ and
$\sim0.7''$ in Bands 3 and 6 respectively. The maximum recoverable scale (MRS) was $\sim10"$ in Band 6 and $\sim13"$ in Band 3 (except for IRAS~19134+2131 with a MRS of $\sim8"$ in Band 3). Since the sources are significantly more compact than the MRS, our results are not affected by resolved out flux. In Table~\ref{Tab:Pos}, we give the coordinates of the unresolved continuum peaks detected towards
each source, because it was noted that for
several sources the catalog positions were inaccurate by up to several
arcseconds. As a result, the rms for the sources with errors in the position is increased due
to the required primary beam corrections. The positional offset for IRAS~18139-1816 was such that the source was not in the ALMA field of view in Band 6 and we thus lack the observations of the $J=2-1$ transitions for this source. It is likely that this IRAS source is not associated with the water fountain OH~12.8-0.9 \cite{Boboltz2007}, whose coordinates match the continuum peak detected using ALMA.  Additionally, due to their
proximity to the Galactic plane, the $^{12}$CO spectra and images of eight sources (IRAS names
15103-5754, 15445-5449, 18043-2116, 18139-1816, 18286-0959,
18450-0148, 18460-0151, and 18596+0315) are
affected by large scale Galactic CO contamination. This can be seen in
the $^{12}$C$^{16}$O,~$J=2-1$ maps towards IRAS~18286-0959 shown in Fig.\ref{fig:12CO_maps}. Since this interstellar emission is spectrally very narrow, it is easily
distinguishable from the spectrally broader
circumstellar emission from the WFs. It is possible to correct for
this effect in the calculations of integrated line fluxes. Additionally, the
lines of the rarer isotopologues on which we base the mass estimates are not
affected by Galactic contamination.

\begin{figure*}[!pt]
 	\centering
 	\includegraphics[angle=-0,width=0.98\textwidth]{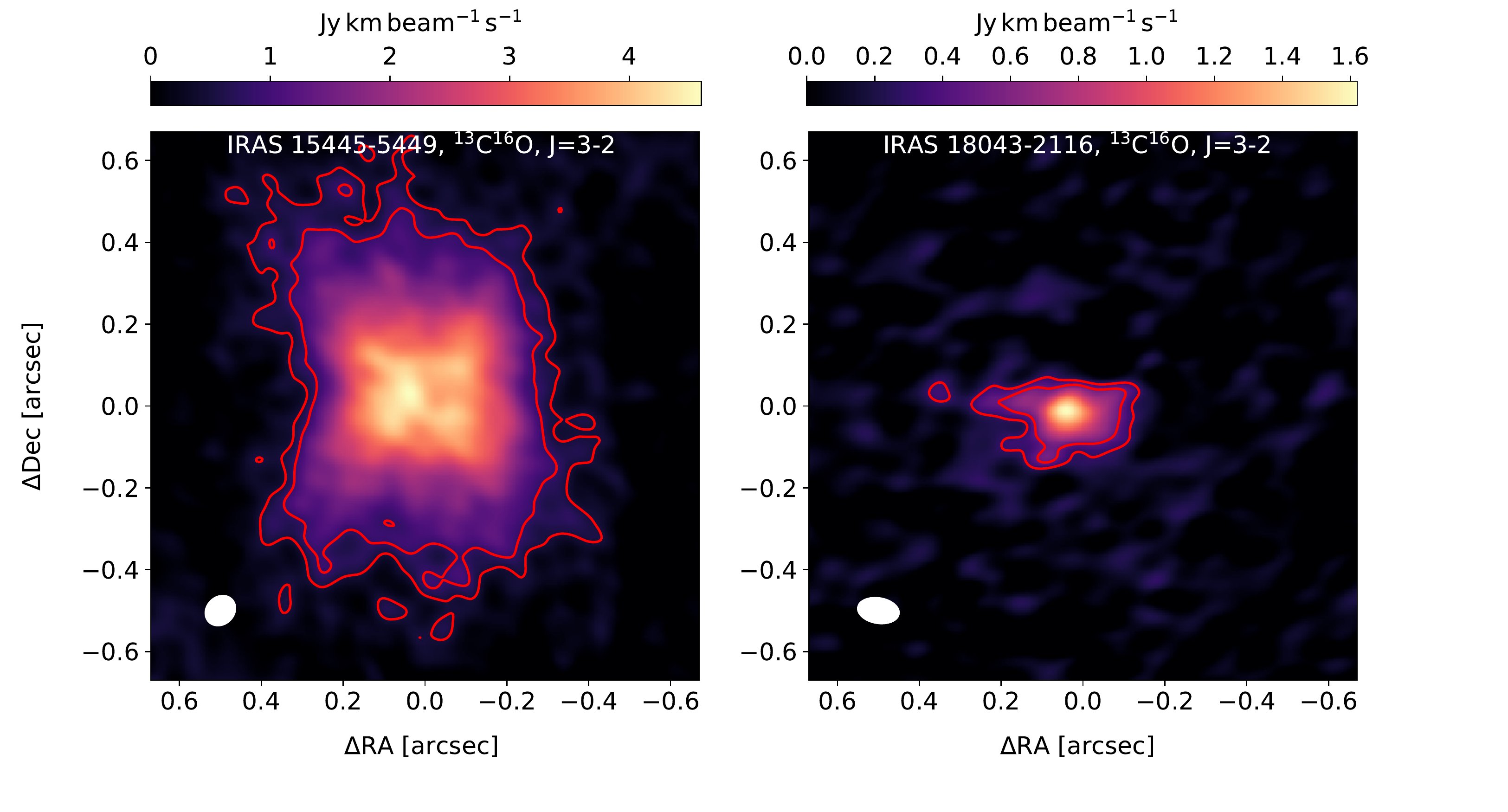}
	\caption{Moment-zero map of the $^{13}$C$^{16}$O,~$J=3-2$ towards IRAS~15445-5449 (left) and IRAS~18043-2116 (right). The dark ellipse shows the beam size. The red contours show the three- and five-$\sigma$ line-flux levels.}
	\label{fig:Andres}
\end{figure*}

In order to estimate the size of the emission region of the $^{12}$C$^{18}$O,~$J=2-1$ line,
we used previous published
observations where available. The estimates were based on the extent of the slow outflow as traced by the main CO isotopologues. For sources IRAS~15103-5754, IRAS~16342-3814, and IRAS~18450-0148, we
used published observations
as discussed below and referenced in Table~\ref{Tab:isotRatios}.
For the WFs IRAS 15445-5449
and IRAS 18043-2116, we used unpublished ALMA observations from project
2016.1.01032.S. The sources in this project were observed at 25
August 2017 (IRAS 18043-2116) and 23 August 2017 (IRAS
15445-5449) in ALMA Band 7 with four spws. These were centered at
319.0, 321.22, 330.58, and 332.5~GHz. The spw at 321.2~GHz had a
bandwidth of 468.75~MHz while the other spws had a bandwidth of
1.875~GHz. All spw used 3840 channels resulting in a resolution of
$\sim1.75$~km~s$^{-1}$ for the three widest spws and
$\sim0.26$~km~s$^{-1}$ for the narrow spw. The data were calibrated
using CASA pipeline version 4.7. Subsequently, {\it uvcontsub} was
used to subtract the continuum and {\it tclean} was used to produce
the spectral line cubes. The $^{13}$CO,~$J=3-2$ line was imaged at
native frequency resolution and Briggs weighing resulting in beam
sizes of $0.076''\times0.065''$ for IRAS~15445-5449 and
$0.095''\times0.06''$ for IRAS~18043-2116. The MRS for these observations was $\sim1"$. The velocity integrated
$^{13}$C$^{16}$O images are shown in Fig.~\ref{fig:Andres}.

We also targeted the strongest of the $^{12}$C$^{18}$O, $J=2-1$ WF sources, IRAS~16342-3814 with the APEX telescope to directly detect the $^{12}$C$^{17}$O, $J=2-1$ transition. The source was observed for 6.5 hours in each transition. The data was processed using the GILDAS package. Both transitions were detected with velocity-integrated fluxes of $2.2\pm0.3$~Jy~km~s$^{-1}$ and $1.36\pm0.28$~Jy~km~s$^{-1}$ for the $^{12}$C$^{18}$O and $^{12}$C$^{17}$O lines respectively.

\subsection*{Distances, luminosities, and nature of water fountains}

The distances were adopted from values available and commonly used in the literature (Table~\ref{Tab:IntLines}). For eight sources, we used the values obtained from integrating the spectral energy distribution and assuming a luminosity in the range expected for (post-)asymptotic giant branch (AGB) stars \cite{Vickers2015}. 
%This can lead to a somewhat circular argument, since the assumption that these are evolved stars is part of the distance estimation method. 
Observations of astrometric or statistical parallaxes for IRAS~18113-2503 \cite{Orosz2019}, IRAS~18286-0959 \cite{Imai2013}, and IRAS~19314+2131 \cite{Imai2007} confirm that these sources have luminosities in the expected range, roughly between a few times $10^{3}$ and $10^{4}$~L$_\odot$. Moreover, kinematic distances for all the sources are consistent with luminosities in this range, as expected for giant evolved stars.

The high luminosity values,  isotopic ratios, and the lifetime of the water fountain phase strongly disfavor interpretations that water fountains are (massive) young-stellar objects or mergers between main-sequence stars (as is expected to be the case of red novae). The uncertainty in the distance for most sources allows the possibility that at least some of them were red-giant-branch stars at the time of the mass-ejection event. In fact, the reported number of sources that appear to have evolved through the CE phase while one of the stars is on the RGB is increasing in recent years \cite{Sahai2017b,Olofsson2019,Kamisnki2021}. Some of these sources might even have evolved through a WF phase. We speculate that sources that interact during the RGB likely corresponds to less than half of the WF objects because most CE interactions are expected to happen during the AGB when stars become the largest. Only accurate distance and initial mass determinations will allow the classification between RGB and AGB to be reliably done for individual sources. The uncertainty on the exact evolutionary phase does not affect in any way our conclusion that these objects are the products of CEE between a giant star and a companion.

\subsection*{Calculation of gas masses{ , average mass-loss rates} and isotopic ratios}

The ALMA observations we present cover the $J=2-1$ transition of $^{12}$C$^{16}$O, $^{13}$C$^{16}$O, 
$^{12}$C$^{18}$O and the $J=1-0$ transition of $^{12}$C$^{16}$O and $^{12}$C$^{17}$O. As shown in Table~\ref{Tab:IntLines}, the lines of $^{12}$C$^{16}$O and $^{13}$C$^{16}$O are detected towards all sources. The only exception is IRAS~18139-1816 because the source position available in the literature did not correspond to the position of the WF source and the Band~6 observations were unsuccessful.
For eight sources, we also detect emission from the two rarer isotopologues ($^{12}$C$^{17}$O and $^{12}$C$^{18}$O), with the exception of $^{12}$C$^{17}$O towards IRAS~15445-5449. We use these observations to derive total gas masses and
the $^{17}$O/$^{18}$O isotopic ratio. We base our calculations mostly on the lines of the rarer isotopologues to avoid difficulties introduced by the high optical depth in the lines of the
main isotopologues. We limit our calculations to the slowly expanding outflow and, hence, integrate the emission
in the $^{12}$C$^{17}$O and $^{12}$C$^{18}$O lines considering an expansion velocity of 20~km/s plus 5~km/s to account for
turbulence in the outflow. For most sources, this is enough to include all observed emission. For IRAS~15103-5754 and
IRAS~18450-0148, we use slightly larger observed values for the velocity of the envelope of
23 and 25~km/s, respectively \cite{Gomez2018,Tafoya2020}.
For IRAS~15445-5449, the equatorial outflow seems to have a significantly larger expansion velocity (65~km/s) than for the other
sources and we adopt this value in our calculations.
For IRAS~16342-3814, the high-velocity outflow seems to contaminate the $^{12}$C$^{18}$O,~$J=2-1$ line. We adopt an
expansion velocity of 20~km/s, and the retrieved flux corresponds to $\sim 65\%$ of the whole line. Hence,
the derived gas mass might be significantly underestimated for this source. The observed $^{12}$C$^{18}$O,~$J=2-1$ lines and the
integration ranges for each source are shown in Fig.~\ref{fig:C18O}.

We assume a Boltzmann distribution of molecules over the rotational levels with an excitation temperature $T_{\rm exc}$.
We assume $T_{\rm exc} = 100$~K based on spatially resolved observations (previously published or available in the ALMA archive) of optically thick lines arising from the low-expansion-velocity gas towards IRAS~15103-5754, IRAS~15445-5449, IRAS~16342-3814, and IRAS~18450-0148. We infer that the $^{12}$C$^{18}$O,~$J=2-1$ and $^{12}$C$^{17}$O,~$J=1-0$ lines for all sources arise mostly from this component of the circumstellar envelope based on the observed line velocity profiles (Fig.~\ref{fig:C18O}). For IRAS~15103-5754, whose torus was directly observed using the $^{12}$C$^{18}$O,~$J=2-1$ line \cite{Gomez2018}, the gas temperatures at the inner and
outer rims of the torus are 180~K and 60~K, respectively, based on optically thick HCO$^+$,~$J=4-3$ emission. The brightness temperature of the $^{12}$C$^{16}$O~$J=2-1$ line averaged over the torus is $\sim 50$~K.
This value is comparable to the colder gas temperatures inferred for the outer regions of the torus in this source, and can be understood considering that the optically thick lines of the main CO isotopologues preferentially trace colder gas in the
outer regions of the envelopes (as well as the other components of the circumstellar envelope).
For IRAS~15445-5449, the optically thick $^{13}$C$^{16}$O,~$J=3-2$ line also
reaches a brightness temperature $\sim 50$~K towards the torus. For IRAS~16342-3814, the $^{12}$C$^{16}$O,~$J=3-2$ lines reaches brightness temperatures $\sim 70$~K. And for IRAS~18450-0148,
the $^{12}$C$^{16}$O,~$J=2-1$ lines has a peak brightness temperature $> 100$~K, although no well-defined
torus is observed for this source, \cite{Tafoya2020}.
% This indicates that the excitation temperatures in the tori (or inner regions) of these sources are similar.
Since the brightness temperatures observed are consistent with the temperature inferred  for the coldest gas at the outer regions of the torus
in IRAS~15103-5754 \cite{Gomez2018}, we consider that an average temperature of $\sim 100$~K as a good approximation for the bulk of the gas of the slowly expanding gas in these sources.
The assumed excitation temperature affects both the gas masses and the isotopic ratios we derive. However, as discussed below, variations of the excitation temperature within acceptable limits do
not affect our conclusions.

For IRAS~16342-3814, we also detected the $^{12}$C$^{17}$O,~$J=2-1$ line using the Atacama Pathfinder Experiment (APEX) in observations carried out in programme ID 0104.F-9310(A). The ratio between the $J=2-1$ and the $J=1-0$ $^{12}$C$^{17}$O lines is $17\pm6$. This translates to an excitation temperature $>35$~K when assuming the lowest line ratio allowed by that one-$\sigma$ uncertainty. The expected ratio for an excitation temperature of 100~K is $\sim 14.5$ further increasing to $\sim 16$ for higher temperatures. Hence, the observations are consistent with our assumed excitation temperature even if they do not provide a very strong constraint.

Assuming optically thin emission, we estimate the total number of $^{12}$C$^{18}$O molecules, $\Xi^{\rm C^{18}O}_{\rm tot}$, from
the integrated flux density of the $^{12}$C$^{18}$O~$J=2-1$ line, $\int S^{\rm C^{18}O_{2-1}}_{\nu} d\nu$, using
%S_nu = N_up*region_area*np.power(d*1.496E11/1000.0,2.0)*A_ul*h*freq/(4.0*np.pi*d*d*3.086E16*3.086E16)/delta_nu
\begin{equation*}
\Xi^{\rm C^{18}O}_{\rm tot} = \frac{4 \pi d^2}{A^{\rm C^{18}O}_{\rm 2-1}~h~\nu^{\rm C^{18}O}_{\rm 2-1}~g^{\rm C^{18}O}_{\rm 2}} e^{(E^{\rm C^{18}O}_{\rm 2}/k T_{\rm exc})} Z^{\rm C^{18}O}(T_{\rm exc}) \int S^{\rm C^{18}O_{2-1}}_{\nu} d\nu.
\end{equation*}
Then, we calculated the hydrogen gas mass using $M^{\rm H_2} = \frac{m^{\rm H_2}~\Xi^{\rm C^{18}O}_{\rm tot}}{X^{^{12}{\rm C}^{18}{\rm O}}}$,
where $m^{\rm H_2}$ is the mass of the H$_2$ molecule and $X^{^{12}{\rm C}^{18}{\rm O}}$ is the fractional $^{12}$C$^{18}$O abundance relative to H$_2$. The total gas masses can be calculated using a helium number abundance $\sim 0.1$ relative to hydrogen after the first-dredge up event (Karakas \& Lugaro, 2016). This implies total gas masses $\sim 40\%$
larger than the hydrogen masses we find. Decreasing the excitation temperature from the assumed value of 100~K by a factor of three decreases the inferred gas mass by a factor $\sim 2.3$, while increasing the assumed excitation temperature by a factor of three increases the derived gas mass by a factor $\sim 2.7$.

{ Choosing a value for $X^{^{12}{\rm C}^{18}{\rm O}}$ is not straight-forward, because the $^{16}$O/$^{18}$O ratio is not known for these sources. The detection of $^{12}$C$^{18}$O emission and the observed $^{12}$C$^{17}$O/$^{12}$C$^{18}$O line ratios imply that WFs have not experienced hot-bottom burning. This is because the surface abundance of $^{18}$O is expected to decrease by orders of magnitude if hot-bottom burning is active \cite{Boothroyd1995}, and the derived gas masses would become unrealistically high in this case. Since the third-dredge up is not expected to affect the oxygen isotopic ratios, the $^{16}$O/$^{18}$O ratio in WFs is set by the enrichment processes before the AGB phase \cite{Lattanzio1997}. The effect of such enrichment is to increase the $^{16}$O/$^{18}$O ratio \cite{Lebzelter2015}. Observations of the $^{16}$O/$^{18}$O ratio in AGB stars reveal a broad range of values. By combining the ratios determined for a total sample of 93 AGB stars \cite{Hinkle2016,Lebzelter2019}, we find a mean value for the $^{16}$O/$^{18}$O ratio of 762 and a median value of 513. In this context, we assume $X^{ ^{12}{\rm C}^{18}{\rm O}} = X^{{\rm C}^{16}{\rm O}}/570 = 5.3\times10^{-7}$, 
obtained considering a $^{12}$C$^{16}$O abundance relative to H$_2$ $X^{^{12}{\rm C}^{16}{\rm O}}=3\times10^{-4}$ and a $^{16}$O/$^{18}$O isotopic ratio of 570. 73\% of the sample of 93 stars considered above has values within a factor of three of 570 (with 12\% having larger values and 15\% lower values).} 

The $^{17}$O/$^{18}$O ratio was derived from the observed line flux ratio also under the assumption of optically thin emission and using
\begin{equation*}
\frac{X^{\rm ^{17}O}}{X^{\rm^{18}O}} = \frac{N^{\rm C^{17}O}}{N^{\rm C^{18}O}} = \frac{A^{\rm C^{18}O}_{\rm 2-1} \nu^{\rm C^{18}O}_{\rm 2-1}}{A^{\rm C^{17}O}_{\rm 1-0} \nu^{\rm C^{17}O}_{\rm 1-0}} \frac{g^{\rm C^{17}O}_{\rm 1} e^{(E^{\rm C^{17}O}_{\rm 1}/k T_{\rm exc})} Z^{\rm C^{17}O}(T_{\rm exc}) \int S^{\rm C^{17}O_{1-0}}_{\nu} d\nu}{g^{\rm C^{18}O}_{2} e^{(E^{\rm C^{18}O}_{\rm 2}/k T_{\rm exc})} Z^{\rm C^{18}O}(T_{\rm exc}) \int S^{\rm C^{18}O_{2-1}}_{\nu} d\nu},
\end{equation*}
where $A^{\rm C^{18}O}_{\rm 2-1}$, $\nu^{\rm C^{18}O}_{\rm 2-1}$, and $S^{\rm C^{18}O_{2-1}}$ are the Einstein-A coefficient, the frequency, and the observed flux density for transition $J=2-1$ of $^{12}$C$^{18}$O,
$g^{\rm C^{18}O}_{2}$ and $E^{\rm C^{18}O}$ are the degeneracy and the excitation energy of state $J=2$ of $^{12}$C$^{18}$O, and $Z^{\rm C^{18}O}(T_{\rm exc})$ is the partition function of the $^{12}$C$^{18}$O molecule at an excitation temperature $T_{\rm exc}$.
The corresponding quantities for transition $J=1-0$ and level $J=1$ of $^{12}$C$^{17}$O are indicated by the C$^{17}$O superscript and the corresponding subscripts.

The derived isotopic ratios have a weak dependence on the assumed excitation temperature. Decreasing the temperature by a factor of two or three causes the derived $\frac{X^{\rm ^{17}O}}{X^{\rm^{18}O}}$ ratio to decrease by less than 10\% and 20\%, respectively. The effect of an
increase in the excitation temperature is even smaller, and changes by factors of two and three increase the estimated $\frac{X^{\rm ^{17}O}}{X^{\rm^{18}O}} $ ratio by less than 6\% and 8\%, respectively.

To test the assumption of optically thin emission, we estimate the optical depth in the $^{12}$C$^{18}$O,~$J=2-1$ line centre using
\begin{equation*}
\tau_\nu = \frac{c^2 A^{\rm C^{18}O}_{\rm 2-1}}{8 \pi \nu^2} \frac{N^{\rm C^{18}O}}{Z^{\rm C^{18}O}(T_{\rm exc})} g^{\rm C^{18}O}_{\rm 2} e^{-\frac{E^{\rm C^{18}O}_{\rm 1}}{kT_{\rm exc}}}(1-e^{\frac{-h \nu}{k T_{\rm exc}}}) \phi,
\end{equation*}
%g_low cancels out from the boltzmann expression
where $\nu = \nu^{\rm C^{18}O}_{\rm 2-1}$, $c$ is the speed of light, $\phi$ is the line shape function (assumed to be box shaped with the observed spectral width), and
$N^{\rm C^{18}O}$ the $^{12}$C$^{18}$O column density calculated using the area of the emission region, $\Sigma$, from $N^{\rm C^{18}O}  = \Xi^{\rm C^{18}O}_{\rm tot}/\Sigma$.
An average mass-loss rate is, then, estimated using the radius of the emission region ($R$), the expansion velocity of the (low-velocity) gas component ($\upsilon_{\rm exp})$, and the calculated gas mass ($M^{\rm H_2}$)
\begin{equation*}
\frac{\Delta M^{\rm H_2}}{\Delta t} = M^{\rm H_2} \frac{\upsilon_{\rm exp}}{R}.
\end{equation*}

The age of the mass-ejection event, $\Delta t$, is a crucial parameter in our calculations of the average mass-loss rates. Given the heterogeneity of the empirical data on the WF sources, a consistent approach for determining $\Delta t$ for all sources is not possible. Hence, we have adopted four different methods to derive the ages, which are shown in Tables~\ref{Tab:1}~and~\ref{Tab:isotRatios}. These are listed below in decreasing order of preference. First, for five sources (IRAS~15103-5754, IRAS~15445-5449, IRAS~16342-3814, IRAS~18043-2116, and IRAS~18450-0148) the extent of the slow outflow was determined directly from spatially resolved observations of lines of CO isotopologues. Second, for IRAS~18460-0151 the age was determined from the extent and expansion velocity of low-velocity H$_2$O masers.
Third, for IRAS~18113-2503 and the sources with no $^{12}$C$^{18}$O detection the age was estimated from the extent of high-velocity H$_2$O maser emission. Fourth, for IRAS~18596+0315,
we use the upper limits on the emission region of $^{12}$C$^{18}$O,~$J=2-1$ derived from the ALMA observations we present.

% For IRAS~18113-2503 (and the sources with no $^{12}$C$^{18}$O detection, see below), we adopted ages obtained from the high-velocity H$_2$O masers.
{ For determining the timescale since the ejection, we assume ballistic expansion of the outflow at the present expansion speed of the low-velocity gas and, hence, the ages of the circumstellar envelopes can be underestimated if acceleration is very slow. Interestingly, the high-velocity outflow in a few sources with observations at higher angular resolution is seen to be decelerating. This is the case for IRAS 16342-3814 \cite{Tafoya2019}, IRAS 18113-2503 \cite{Orosz2019}, and IRAS 18450-0148 \cite{Tafoya2020}. This does not imply, however, that the same is true for the slow component of the outflow. Nonetheless, the velocities we measure from the $^{12}$C$^{18}$O line are relative slow. Even if we consider the weakest possible constant acceleration from 0~km/s to 20~km/s, the derived ages would still be only $\sim500$ years for an envelope of 1000 au in radius. This hypothetical weak acceleration is a lower limit and is not supported by the observations, because sources with envelopes of different sizes show very similar expansion velocities. Hence, a relatively strong acceleration seems necessary. It is also not obvious what the source of a weak acceleration would be. The momentum carried by the radiation field of a star with 5000~L$_\odot$ only matches the linear momentum of a 0.1 solar mass of gas expanding at 20~km/s after 20 thousand years. Given that the crossing time to reach the observed envelope sizes (or the upper limits) is much lower than this even for very weak constant acceleration, we conclude that radiation pressure probably does not contribute significantly to the acceleration of the outflow. This is similar to the conclusion by \cite{Bujarrabal2001} based on observations of the fast outflows in post-AGB stars. In this context, a sudden acceleration following an episodic ejection of gas seems the most likely scenario for WFs. In conclusion, the uncertainty on the derived mass-loss rates by assuming ballistic expansion is probably relatively small (although difficult to estimate precisely) and does not necessarily imply that the age of the envelope is underestimated.}

{ Another important consideration is that }
the age from the fast outflow might not be comparable to that of the slow outflow for a given source. To investigate this, we compare these two ages for the sources for which both have been determined. Observations of the fast CO outflows reveal ages between 70-100~yr for IRAS~16342-3814 \cite{Sahai2017} and $\sim60$~yr for IRAS~18450-0148  \cite{Tafoya2020}. These values are $\sim 40\%$ lower than the values we infer for the slow-outflow component, $\sim 154$~yr and $\sim 90$~yr. For IRAS~15103-5754, the ages of the fast and slow outflows are consistent \cite{Gomez2018}, and also consistent with the age we use. For IRAS~18043-2116, the age $\gtrsim 150$~yr based on observations of H$_2$O masers \cite{Walsh2009} is also consistent with the age derived by us for the slow-component of the outflow { (195~yr)}. The only source with a significant difference between the ages of the slow outflow, $\sim 150$~yr, and the fast outflow, $\sim 20$~yr, is IRAS~18460-0151 \cite{Imai2013b}. This object has the smallest value of the fast-outflow age from the whole WF sample and is also the only source with a slowly expanding envelope spatially resolved in the $^{12}$C$^{16}$O,~$J=2-1$ maps. Hence, it is probably not representative of the typical WF. In our calculations for this source, we adopt the age of the slow-outflow component determined from low-velocity H$_2$O masers, $\sim 150$~years \cite{Imai2013b}.
% This is in agreement of estimates for other evolved stars with fast and slow outflows which show the timescales of these two components differs at most by a factor of three \cite{Huggins2007}.

Overall, the ages from the fast outflows prove to be a good approximation for the time of the ejection events. Most importantly for our purposes, $\sim 200$~yr seems to be a clear upper limit for the ages of WFs, since ages significantly larger have not been reported by any method for any WF. Hence, we adopt the ages from the fast outflows in our calculations when the other two preferred methods cannot be used. For the sources with $^{12}$C$^{18}$O detection, this is only necessary for IRAS~18113-2503. The average mass-loss rate value derived for this source increase by more than an order of magnitude when considering the age from the fast outflow (Table~\ref{Tab:isotRatios}), rather than the upper limits from the ALMA maps. The average mass-loss rate is at least one order of magnitude higher than expected for a star with initial mass $< 4$~M$_\odot$ independently of which of the two ages we adopt for this source. The smaller emission region implied by the shorter age, causes the $^{12}$C$^{18}$O,~$J=2-1$ line to reach an optical depth comparable to one (Table~\ref{Tab:isotRatios}). This would make our approach for deriving the gas masses not ideal (and would indicate the gas masses are potentially larger than we find). A direct determination of the $^{12}$C$^{18}$O,~$J=2-1$ emission region in this source will show whether the ages from the fast outflows can be translated into mass-ejection timescales for the slowly expanding outflow component. %For these two sources, even the lower limits on the average mass-loss rates obtained from the poor constraints on the size of the $^{12}$C$^{18}$O~$J=2-1$ emission region imply very large values.
% We find that the $^{12}$C$^{18}$O,~$J=2-1$ line
% is indeed expected to be optically thin towards all these sources (Table~\ref{Tab:isotRatios}).

%tau_nu = c*c/(8.0*np.pi*np.power(freq,2.0))*(g_up/g_low)*N*f_low*A_ul*(1-np.exp(-h*freq/(k*Texc)))*phi_center

The derived average hydrogen mass-loss rates are proportional to the distances to the sources, inversely proportional to $X^{^{12}{\rm C}^{18}{\rm O}}$, and depend on $T_{\rm exc}$ as discussed above.
For most sources, the distances are kinematic distances based on the source velocity, and the uncertainty is relatively large.
{ Considering uncertainties of a factor of three on the distances, $T_{\rm exc}$ and $X^{^{12}{\rm C}^{18}{\rm O}}$, we estimate the derived average mass-loss rates to be uncertain by
a factor of five.} Even taking this uncertainty estimate into account, the average hydrogen mass-loss rates derived
are too high by more than an order of magnitude for processes other than CEE.

For the sources with no $^{12}$C$^{18}$O detection, we calculated upper limits to the hydrogen masses using the same procedure as detailed above and the upper limit on the line fluxes derived from the observations, which is the three-$\sigma$ flux uncertainty over the expected line width. To further constrain the hydrogen masses, we calculated lower limits using the $^{13}$C$^{16}$O,~$J=2-1$ lines. To be conservative, we used a low excitation temperature ($40$~K) and a low $^{12}$C/$^{13}$C isotopic ratio
equal to five. We find lower limits on the hydrogen masses between 0.01 and 0.03~$M_\odot$. This shows
that the gas masses in the circumstellar environment of these sources are still relatively large. Our conservative assumptions and the probable high optical depth in the $^{13}$C$^{16}$O line suggest that the actual hydrogen masses might be considerably larger than these lower limits. Considering
the ages obtained from literature, these imply lower limits on the average mass-loss rates between $10^{-4}$ and $4\times 10^{-4}~M_\odot$/yr, which are also larger than expected for stars with initial masses $< 4~M_\odot$.
%The only exception to this is IRAS~18286-0959, for which our calculations imply an average mass-loss rate between $3\times10^{-5}$ and $2\times10^{-4}~M_\odot$/yr, which is close to what is expected for stars at the tip of the AGB. 

%\begin{figure*}
%	\centering
%	\includegraphics[angle=-0,scale=0.85]{Mass_vs_isotope_C17O_C18O_ratio.pdf}
%	\caption{C$^{17}$O/C$^{18}$O ratio as a function of the molecular mass. The Horizontal dotted line represents the solar $^{17}$O/$^{18}$O ratio.}
%	\label{Mass_vs_isotope_ratio_C17O_C18O}%
%\end{figure*}
%
%
%\begin{figure*}
%	\centering
%	\includegraphics[angle=-0,scale=0.85]{Mass_vs_isotope_C16O_C17O_ratio.pdf}
%	\caption{C$^{16}$O/C$^{17}$O ratio as a function of the molecular mass. The Horizontal dotted line represents the solar $^{16}$O/$^{17}$O ratio.}
%	\label{Mass_vs_isotope_ratio_C16O_C17O}%
%\end{figure*}

%\begin{figure*}
%	\centering
%	\includegraphics[angle=-0,scale=0.85]{W12CO_vs_WC18O.pdf}
%	\caption{Velocity-integrated flux of $^{12}$CO vs. C$^{18}$O}
%	\label{Mass_vs_isotope_ratio_C16O_C17O_2}%
%\end{figure*}
%
%\begin{figure*}
%	\centering
%	\includegraphics[angle=-0,scale=0.85]{W12CO_vs_W13CO.pdf}
%	\caption{Velocity-integrated flux of $^{12}$CO vs. $^{13}$CO}
%	\label{Mass_vs_isotope_ratio_C16O_C17O_3}%
%\end{figure*}

\subsection*{Isotopic ratios derived for the water fountains in context}

The $^{17}$O/$^{18}$O isotopic ratios we find imply initial stellar masses $\lesssim 4$~M$_\odot$. The characteristics of six out of the eight sources with rarer-isotopologue detection,
however, are inconsistent with theory for stars at the tip of the AGB. The inferred initial mass range
%$^{12}$C/$^{13}$C, $^{16}$O/$^{17}$O, and $^{16}$O/$^{18}$O ratios might be compatible if the optical 
%depths in the main isotopologues are large enough, the
implies that these six objects should be carbon rich by the end of their evolution for initial solar composition, but all WFs are oxygen rich.
For an initial solar isotopic ratio,
the lower limit on the $^{17}$O/$^{18}$O ratio for becoming a carbon star at the end of the AGB (or experiencing hot-bottom burning)
is $\sim 1.0$, $\sim0.8$, and $\sim1.4$ for metallicity $Z=0.007$, 0.014, and 0.03 \cite{Karakas2016}. All six sources have $^{17}$O/$^{18}$O at least one $\sigma$ larger than these limits.
For three sources with observed $^{17}$O/$^{18}$O ratios larger than three, our results are even more constraining, because the ratios are too large for models with $Z=0.03$ and initial masses $< 4$~M$_\odot$ at the tip of the AGB, which predict ratios smaller than two. Hence, their ratios are too high for them to be metal rich at the tip of the AGB, or metal poor and not carbon-rich stars.

Observations of oxygen isotopic ratios in stars in the AGB \cite{Hinkle2016,DeNutte2017,Lebzelter2019} reveal values that are mostly in agreement with expectations from theoretical models.
For some sources, the rarer isotopes are found to be more abundant than expected relatively to the main one, which is usually interpreted a sign of initial abundances that differ from solar \cite{Lebzelter2015}.
In our calculations, we assume a value of 570 for the $^{16}$O/$^{18}$O ratio, which is about the lowest value expected for AGB stars with initial solar composition after the first dredge-up.
If the WF systems have enhanced $^{18}$O abundances, the derived masses and $^{17}$O
abundances will be overestimated by the same factor. This would make the inferred $^{17}$O abundance high compared to models for four sources in our sample.
Values of the $^{16}$O/$^{18}$O ratio lower than we assume by more than a factor of { three} are unlikely based on the observations of other AGB stars and on the chemical evolution of the Galaxy \cite{Prantzos1996}.
%Should this go here or should I move (part of it) to the Appendix?

%The derived $^{17}$O/$^{18}$O is also too large for three sources. 
%We discuss the following possibilities to explain this discrepancy:

Three sources (IRAS~18043-2116, IRAS~18113-2503, and IRAS~18596+0315) have $^{17}$O/$^{18}$O ratios that are larger than three, while the maximum expected from models for solar metallicity after the first dredge-up event is $\sim 3.2$. Possible explanations for this include { contamination from a more massive companion}, excitation temperature, an initial chemical composition that differs from solar, optical depth effects, or a combination of these. However, this is at most at the two-$\sigma$ level and our data does not allow us to come to a conclusion on this topic. Nonetheless, our conclusions on the nature of the WFs are not dependent on the exact values of the $^{17}$O/$^{18}$O isotopic ratios.

{ We also consider whether companions could significantly affect the abundance of $^{17}$O and $^{18}$O in the circumstellar gas, because this could hamper the interpretation of isotopic ratios based on evolutionary models for single stars. While contamination from companions can potentially affect the derived circumstellar masses and isotopic ratios for WFs, it would obviously not change our main conclusion that the WF phase is the consequence of binary interactions. We identify two possible scenarios: contamination from a more massive companion preceding the current CE phase and contamination from a main-sequence companion during the current CE phase. Regarding the first case, the $^{18}$O surface abundance is only expected to decrease by a relatively small factor, $\lesssim 2$ \cite{Lebzelter2015}, during the evolution of low- and intermediate-mass stars up to the AGB. During AGB evolution the surface abundance of $^{18}$O is either not modified or sharply decreases if hot-bottom burning is active.  Hence, any contamination from a more massive companion evolving through the AGB would not have increased the $^{18}$O surface abundance. Contamination from a supernova could substantially increase the $^{18}$O surface abundance, but this is not a statistically plausible scenario for WFs. Contamination from a more massive companion, which  experienced hot-bottom burning and would now be a white dwarf, would increase the $^{17}$O/$^{18}$O ratio in the envelope of the now giant star, since the accreted gas would be very $^{18}$O poor. This could be an explanation for the high values of the $^{17}$O/$^{18}$O ratio observed towards a few of the WFs. The increase in $^{17}$O would imply a decrease in $^{18}$O abundance in this case, and potentially larger gas masses than we derive. However, as mentioned above, we are unable to come to a conclusion on this matter based on the data at hand. Regarding a main-sequence companion, we find it unlikely that the isotopic ratios of the circumstellar gas would be affected in this case. First, a main-sequence companion is not expected to contribute a significant fraction of mass to the circumstellar gas \cite{Taam2000}. Second, a main sequence companion would have an $^{18}$O abundance less than a factor of two higher than the giant star \cite{Lebzelter2015} and a lower $^{17}$O/$^{18}$O ratio. Hence, even if a main-sequence companion would eject a significant amount of gas in a CE interaction, this could not explain the high $^{17}$O/$^{18}$O observed towards some sources, nor would it have a significant effect on the $^{18}$O abundance and on the circumstellar gas masses derived by us. In conclusion, the main effect from companions on the observed oxygen isotopic ratios is that accretion of gas from more massive companions could have decreased the $^{18}$O abundance in the now giant star, but such an evolutionary path has probably only been experienced by a small number of WFs.}

\subsection*{Water-fountain and common-envelope systems formation rates}

Currently 15 WF sources have been identified throughout
the Milky Way. Their strong H$_2$O maser emission allows them
to be detected throughout the entire Galaxy, but it is difficult to estimate the completeness of the
WF sample. With the
exception of the H$_2$O Southern Galactic Plane Survey \cite{Walsh2011}, there have been few blind surveys for H$_2$O masers. Nonetheless, surveys
based on color criteria have covered a large fraction of the
heavily obscured evolved stars \cite{Engels1996}. 
%Based on HOPS detections of only one previously unknown WF source (Walsh et al. 2009) we estimate
%the WF sample to be up to $\sim 75\%$ complete. 
Considering 15 WFs are currently known and taking $\sim200$~yr as the maximum time during
which we can identify WF sources , this leads to an
estimated WF source rate of $>0.075$~yr$^{-1}$.

Here we compare this rate with the expected rate of recent CE
events. We perform a simple analysis using the initial mass function \cite{Kroupa2001}
scaled to the recently determined star formation history of the Milky
Way \cite{Ruiz-Lara2020}. This relative star formation history is further
scaled to the average recent Galactic star formation rate of
$1.65\pm0.19$~M$_\odot$~yr$^{-1}$ \cite{Licquia2015}. Subsequently we determine the number of stars that have been
able to evolve to the AGB phase within less than a Hubble time
($M>0.95~M_\odot$). We then apply the criteria for CE envelope
evolution at the late-RGB or early-AGB (a mass ratio~$>0.1$ and
period~$<10$~yr) and the corresponding close binary fractions \cite{Moe2015}. This leads to an estimate of the CE rate
for all low mass stars ($<8$~M$_\odot$) of
$0.15\pm0.05$~yr$^{-1}$. This is consistent with previously determined CE rates considering approximately half the Galactic CE events when at least one of the stars has evolved from the main sequence \cite{Kochanek2014}. The rate can also be compared to a total rate of
$0.706\pm0.233$~yr$^{-1}$ for stars evolving through the AGB. The
rates for specific mass bins is presented in Table~\ref{Tab:3}. From this it is
clear that the majority of CE events will occur for stars
$<2$~M$_\odot$, consistent with our derived initial masses for several
of the WF sources. The WF rate thus constitutes a significant fraction of all recent CE
events. Even if the WFs are not the result of close binary
interaction, the rate calculations still indicate that the
progenitors of most of these sources should be $<2$~M$_\odot$. This would
require a hitherto unknown mechanism that is able to eject of order
half the stellar envelope within a few hundred years without interaction with
a close companion.

The interaction with a possible massive planet could also be considered a form of CEE \cite{Villaver2009} and, hence, our calculations might underestimate the rate of occurrence of CE systems.
The fraction of K, G, and F main-sequence stars with planets with masses between 0.3 and 15 Jupiter masses and
orbits $\lesssim 10$~AU is $\sim 14\%$ \cite{Cumming2008}. Therefore, even if such planets are 100\%
efficient in triggering CEE, the rate of K, G, and F stars undergoing CEE in the Galaxy would increase by $< 50\%$, and
the WF phase would still represent a major evolutionary path in CEE.

% \section*{Supplementary text}

\subsection*{Carbon and oxygen isotopic ratios in AGB stars}

In star with masses between 0.8 and 8~$M_\odot$, the change in composition caused by nuclear burning 
is mostly confined to the stellar interior, where conditions allow nuclear burning to take place. The 
surface abundances are only modified from their initial values when convective streams reach these 
inner regions and bring the nuclear burning products to the upper layers. This mixing process is 
referred to as a dredge-up and three distinct episodes can take place during the lives of stars in 
this mass range, one before each of the two ascents on the giant branch and one during the upper 
AGB phase \cite{Herwig2005}. The surface abundances at a given time can be 
calculated using stellar evolution codes, and a large number of studies have contributed to our present 
understanding of surface enrichment up to and during the AGB phase based on such calculations 
\cite{ElEid1994,Lattanzio1997,Boothroyd1999,Busso2010,Palmerini2011,Karakas2014,Straniero2017,Abia2017}. In this study, we focus on the surface abundances of isotopes of oxygen and carbon, which 
 are mainly affected by the first and third dredge-up events. Since the nuclear burning products
  and the efficiency of the different dredge-up episodes depend strongly on the initial stellar mass, 
  surface isotopic ratios (and in particular the oxygen and carbon isotopic ratios) can be, and have 
  been, used to infer the initial stellar mass \cite{Tsuji2007,Ramstedt2014,Hinkle2016,DeNutte2017,Danilovich2017,Lebzelter2019}.

In a gas with solar abundances the isotopic ratios are $^{12}$C/$^{13}$C = 90, $^{16}$O/$^{17}$O = 2600, and 
$^{16}$O/$^{18}$O = 500 \cite{Asplund2009}. The first dredge-up happens when stars ascend 
the RGB and it causes the $^{12}$C/$^{13}$C isotopic ratio to decrease to values between 20 and 
30 and the $^{17}$O abundance to increase, while the abundances of $^{16}$O and $^{18}$O remain mostly 
unaltered \cite{Abia2017}. The change in $^{17}$O abundance is found to be 
particularly sensitive to the initial stellar mass, with a sharp decline of the $^{16}$O/$^{17}$O with increasing 
stellar mass between 1 and 2~$M_\odot$ and a more gradual increase of the ratio for increasing stellar 
masses above 2~$M_\odot$ \cite{Straniero2017}. The third dredge-up episode is 
actually a series of events that happens during the late part of the AGB phase and enriches the surface 
with $^{12}$C produced from helium nuclear burning. The effect can be large enough to create carbon 
stars, in which the surface abundance of carbon is larger than that of oxygen. Only stars more massive 
than ~1.5 $M_\odot$ will go through enough events in the third dredge-up to become carbon stars. Stars 
more massive than about 4 $M_\odot$ experience hot bottom burning (HBB), which means that the CNO 
cycle operates at the base of the convective envelope and causes carbon and oxygen to be converted 
mainly into $^{14}$N \cite{Boothroyd1993}. For stars that do not experience HBB, the 
oxygen abundances remain virtually unchanged during the third dredge-up, while the $^{12}$C/$^{13}$C ratio 
is expected to increase as the $^{12}$C abundance increases. If HBB operates, the $^{17}$O abundance 
slightly increases \cite{Lattanzio1997}, the $^{12}$C/$^{13}$C ratio decreases (down to 
the CN-cycle equilibrium value of ~3 for the more massive AGB stars \cite{Doherty2014}), and the $^{18}$O abundance decreases sharply by a few orders of magnitude 
\cite{Boothroyd1995}.

Observations of isotopic ratios show general agreement with these theoretical expectations 
%(Ramstedt \& Olofsson 2014, A\&A 566, A145; Hinkle et al. 2016, ApJ, 825, 38).
For instance, 
large values of the $^{16}$O/$^{18}$O ratio have been found for extreme OH/IR stars, suggesting these 
sources have experienced hot-bottom burning \cite{Justtanont2015}. Observations 
also reveal discrepancies, including low $^{12}$C/$^{13}$C ratios in RGB stars $\sim 10$ \cite{Tsuji2007} 
usually explained introducing extra mixing processes \cite{Boothroyd1999,Nollet2003}, low $^{12}$C/$^{13}$C ratios in J-type carbon stars $\sim 3$ 
\cite{Abia2003} potentially explained through extrinsic enrichment; and low $^{16}$O/$^{18}$O 
ratios in some sources, which might be caused by initial abundances that differ from solar \cite{Lebzelter2015}. The extra-mixing processes invoked to explain low $^{12}$C/$^{13}$C 
ratios in red giant stars are not expected to affect the oxygen isotopic ratios \cite{Hinkle2016}. The initial values of the isotopic ratios vary with metallicity, with the relative abundance 
of the rarer isotopes expected to increase with increasing metallicity \cite{Kobayashi2011}.

% \afterpage{\clearpage}

% \afterpage{\clearpage}

\pagebreak
% \afterpage{\clearpage}

\begin{table*}[!pht]
\begin{threeparttable}
	\caption{Velocity-integrated flux of observed lines of isotopologues of CO}             % title of Table
	\label{Tab:IntLines}      % is used to refer this table in the text
 	\begin{center}
		\begin{tabular}{ c c c c c c}
			\hline\hline
			Source name & $d$ & $W_{^{12}{\rm CO}(2-1)}$ & $W_{^{13}{\rm CO}(2-1)}$  & $W_{{\rm C}^{18}{\rm O}(2-1)}$  & $W_{{\rm C}^{17}{\rm O}(1-0)}$  \\
			& (kpc) & (Jy~km~s$^{-1}$)& (Jy~km~s$^{-1}$) & (Jy~km~s$^{-1}$) & (Jy~km~s$^{-1}$)  \\
			\hline 
			IRAS 15103$-$5754 & $3.4^{\rm a}$ & $223.0\pm0.1$ & $37.2\pm0.1$ & $2.16\pm0.03$ & $0.32\pm0.02$\\ 
			IRAS 15445$-$5449 & $4.4^{\rm a}$ & $119.6\pm0.1$ & $41.4\pm0.1$ & $1.05 \pm 0.04$ & $<0.07$\\ 
			IRAS 16342$-$3814 & $2.2^{\rm b}$ & $158.5\pm0.2$ & $59.0\pm0.2$ &$2.19 \pm 0.03$&$0.08\pm0.02$\\
			IRAS 16552$-$3050 & $8.4^{\rm a}$ & $2.7\pm0.1$ & $1.2\pm0.1$ & \textit{ND} & \textit{ND} \\ 
			IRAS 17291$-$2147 & $11.7^{\rm c}$ & $2.8\pm0.1$ & $1.3\pm0.1$ & \textit{ND} & \textit{ND} \\ 
			IRAS 18043$-$2116 & $8.2^{\rm a}$ & $11.4\pm0.1$ & $4.6\pm0.1$ & $0.34 \pm 0.02$ & $0.09\pm0.02$\\
			IRAS 18113$-$2503 & $12.0^{\rm d}$ & $17.0 \pm 0.1$ & $4.2 \pm 0.1$ & $0.22\pm0.02$ & $0.09\pm0.02$\\
			IRAS 18139$-$1816 & $5.2^{\rm a}$ & \textit{NObs} & \textit{NObs} & \textit{NObs} & \textit{\textit{ND}} \\
			IRAS 18286$-$0959 & $3.6^{\rm e}$ & $8.6\pm0.1$ & $1.7\pm0.1$ & \textit{ND} & \textit{ND} \\ 
			IRAS 18450$-$0148 & $2.2^{\rm f}$ & $66.6\pm0.2$ & $22.0\pm0.2$& $1.80\pm0.06$ & $0.24\pm0.02$\\
			IRAS 18455+0448 & $11.2^{\rm g}$ & $3.3\pm0.1$ & $1.2\pm0.1$ & \textit{ND} & \textit{ND} \\  
			IRAS 18460$-$0151 & $3.1^{\rm a}$ & $57.3\pm0.1$ & $15.2\pm0.1$& $1.02\pm0.03$ & $0.13\pm0.02$\\
			IRAS 18596+0315 & $8.3^{\rm a}$ & $25.9\pm0.1$ & $6.8\pm0.1$ & $0.43\pm0.02$ & $0.12\pm0.02$\\
			IRAS 19134+2131 & $8.0^{\rm h}$ & $3.9\pm0.1$ & $1.4\pm0.1$& \textit{ND} & \textit{ND} \\
			IRAS 19190+1102 & $8.9^{\rm a}$ & $9.3\pm0.1$ & $3.3\pm0.1$& \textit{ND} & \textit{ND} \\  
			\hline 
		\end{tabular}
		\begin{tablenotes}
      \small
      \item \textit{ND} - not detected, \textit{NObs} - not observed. References: a - \cite{Vickers2015}, b - \cite{Sahai2017}, c - kinematic distance estimated from \cite{Reid2019}, d - \cite{Orosz2018,Orosz2019}, e - \cite{Imai2013}, f - \cite{Tafoya2020}, g - \cite{Vlemmings2014}, h - \cite{Imai2007}.
      \end{tablenotes}
	\end{center}
	\end{threeparttable}
\end{table*}
% \afterpage{\clearpage}

\definecolor{Gray}{gray}{0.9}

\begin{table*}[!pht]
\begin{center}
\begin{threeparttable}
\caption{Physical parameters of the water fountain sources derived from $^{12}$C$^{18}$O and $^{13}$C$^{16}$O emission}             % title of Table
\label{Tab:isotRatios}      % is used to refer this table in the text
	\begin{tabular}{ c c @{}c @{ }c @{ }c @{ }c @{}c}
		\hline\hline
		IRAS name & $M^{\rm H_2}$ & $R$ & $\upsilon_{\rm exp}$ & $\Delta t$ & $\Delta M^{\rm H_2}/\Delta t$ & $\tau^{\rm ^{12}C^{18}O}_{J=2-1}$\\
		  & ($M_{\odot}$) & (au) & (km/s) & (yr) & ($10^{-3}~M_\odot~{\rm yr}^{-1}$) \\
		\hline 
% 		IRAS 15103 &3.4&0.85 & $1000^{\rm a}$  & 21 & 230 & $4\times 10^{-3}$ & 0.02 \\ 
% 		IRAS 15445 &5.0&0.75 & $1000^{\rm b}$  & 32 & 150 & $1 \times 10^{-2}$ & 0.04 \\ 
% 		IRAS 16342 &2.2&0.38 & $880^{\rm c}$  & 35 & 120 & $6 \times 10^{-3}$ & 0.02 \\
% 		IRAS 18043 &8.2&0.61 & $410^{\rm b}$ & 13 & 200 & $6 \times 10^{-3}$ & 0.7 \\ 
% 		IRAS 18113 &10.0&0.60 & $<4600$ & 13 & $<1750~(40^{\rm e})$ & $> 7 \times 10^{-4}$ & $>0.004$ \\ 
%         \rowcolor{Gray}
% 		IRAS 18113 &10.0&0.60 & $220$ & 13 & 40 & $1 \times 10^{-2}$ & $1.75$ \\ 
% 		IRAS 18450 &2.2&0.27 & $250^{\rm d}$ & 9 & 130 & $2 \times 10^{-3}$ & 0.2 \\ 
% 		IRAS 18460 &2.0&0.11 & $<1100$ & 10 & $<750~(20^{\rm f})$ & $>3 \times 10^{-4}$ & $>0.06$ \\
%         \rowcolor{Gray}
% 		IRAS 18460 &2.0&0.11 & $80$ & 10 & 20 & $5 \times 10^{-3}$ & $3.2$ \\
% 		IRAS 18596 &3.0&0.10 & $<1500$ & 20 & $<260$ & $>8 \times 10^{-4}$ & $>0.007$\\ 
		
% 		IRAS 15103$-$5754 &3.4  & $0.68$ & $1000^{\rm a}$ & $23$ & $206$ & $3.3$\\ 
% 		IRAS 15445$-$5449 &5.0  & $0.71$ & $1000^{\rm b}$ & $65$ & $75$ & $10$\\ 
% 		IRAS 16342$-$3814 &2.2  & $0.29$ & $650^{\rm c}$ & $20$ & $150$ & $2.0$\\
% 		IRAS 18043$-$2116 &8.2  & $0.62$ & $410^{\rm b}$ & $20$ & $100$ & $6.4$ \\ 
% 		IRAS 18113$-$2503&10.0  & $0.60$ & $<4600~(220)$ & $20$ & $<1100~(50)$ & $>0.5~(10)$\\ 
% 		IRAS 18450$-$0148 &2.2  & $0.24$ & $500$ & $1.8$ & $25$ & $2.5$\\ 
% 		IRAS 18460$-$0151 &2.0  & $0.11$ & $<1500~(80)$ & $20$ & $<360~(20)$ & $> 0.3~(6)$\\
% 		IRAS 18596$+$0315 &3.0  & $0.11$ & $<1100$ & $20$ & $<260$ & $> 0.4$\\ 

		15103$-$5754 & $0.68$ & $1000^{\rm a}$ & $23^{\rm a}$ & $206$ & $3.3$ & $0.05$ \\ 
		15445$-$5449 & $0.55$ & $1500^{\rm b}$ & $65^{\rm b}$ & $110$ & $5$ & $0.003$\\ 
		16342$-$3814 & $0.29$ & $650^{\rm c}$ & $20$ & $154$ & $2.0$ & $0.05$ \\
		18043$-$2116 & $0.62$ & $820^{\rm b}$ & $20$ & $195$ & $3.2$ & $0.02$\\ 
		18113$-$2503 & $0.87$ & $220(<6800^{\rm d})$ & $20$ & $65^{\rm e}(<1600^{\rm d})$ & $14(>0.5^{\rm d})$ & $0.9(> 0.01^{\rm d})$ \\ 
		18450$-$0148 & $0.24$ & $500^{\rm f}$ & $25^{\rm f}$ & $95$ & $2.5$ & $0.06$ \\ 
		18460$-$0151 & $0.27$ & $80(<1500^{\rm d})$ & $20$ & $150^{\rm g}(<360^{\rm d})$ & $2(> 0.8^{\rm d})$ & $0.07(> 0.01^{\rm d})$ \\
		18596$+$0315 & $0.80$ & $<4100^{\rm d}$ & $20$ & $<975^{\rm d}$ & $> 0.8^{\rm d}$ & $>0.004^{\rm d}$ \\ 
		\hline
        % 16552$-$3050 & $\lesssim0.12$ & $\sim 420$ & $20$ & $\sim 100$ & $\lesssim 1.2$ & $<0.1$ \\
        16552$-$3050 & $0.02~\text{--}~0.12$ & & & $\sim 100^{\rm h}$ & $0.2~\text{--}~1.2$ &  \\
        % 17291$-$2147 & $\lesssim0.25$ & $\sim 1050$ & $20$ & $\sim 250$ & $\lesssim1.0$ & $<0.04$ \\
        17291$-$2147 & $0.04~\text{--}~0.22$ & & & $\sim 250^{\rm i}$ & $0.2~\text{--}~0.9$ &  \\
        % 18286$-$0959 & $\lesssim0.02$ & $\sim 430$ & $20$ & $\sim 30$ & $\lesssim0.7$ & $<0.02$ \\
        % 18139$-$1816 &  &  &  &  &  &  \\
        18286$-$0959 & $0.005~\text{--}~0.02$ &  &  & $\sim 30^{\rm j}$ & $0.1~\text{--}~0.7$ &  \\
        % 18455+0448 & $\lesssim 0.21$ & $\sim 315$ & $20$ & $\sim 75^{\rm g}$ & $\lesssim3.0$ & $<0.3$\\
        18455+0448 & $0.03~\text{--}~0.21$ &  &  & $\sim 75^{\rm k}$ & $0.5~\text{--}~3.0$ & \\
        % 19134+2131 & $\lesssim 0.12$ & & $20$ & $\sim 55$ & $\lesssim 2.2$ &  \\
        19134+2131 & $0.02~\text{--}~0.1$ & &  & $\sim 55^{\rm l}$ & $0.4~\text{--}~1.8$ &  \\
        % 19190+1102 & $\lesssim 0.13$ & $\sim 320$ & $20$ & $\sim 75$ &  
        19190+1102 & $0.06~\text{--}~0.13$ & & & $\sim 75^{\rm m}$ & $0.8~\text{--}~1.7$ &  \\
		\hline 
	\end{tabular}
\begin{tablenotes}
      \small
    %   \item Note: the rows with gray background show values considering the dynamical ages for
    %   IRAS~18113-2503 and IRAS~18460-0151.
      \item References: a - \cite{Gomez2018}, b - Fig.~\ref{fig:Andres} (and Perez-Sanchez et al., in prep), c - \cite{Sahai2017}, d - ALMA observations presented here; e - \cite{Orosz2019}, f - \cite{Tafoya2020}, g - \cite{Imai2013b}, h - estimated from \cite{Suarez2008}, i - estimated from \cite{Gomez2017}, j - \cite{Imai2020}, k - \cite{Vlemmings2014}, l - \cite{Imai2007}, m - \cite{Day2010}.
    \end{tablenotes}%
\end{threeparttable}
\end{center}
\end{table*}
% \afterpage{\clearpage}

\begin{table*}[!p]
% \begin{threeparttable}
	\caption{Positions of the unresolved continuum emission peaks detected in the ALMA observations.}             % title of Table
	\label{Tab:Pos}      % is used to refer this table in the text
 	\begin{center}
		\begin{tabular}{c c c}
			Source & RA & Dec \\
			& (hh:mm:ss) & (dd:mm:ss)\\
			\hline
			
IRAS 15103$-$5754 & 15:14:18.39 & -58:05:20.4 \\
IRAS 15445$-$5449 & 15:48:19.39 & -54:58:20.1 \\
IRAS 16342$-$3814 & 16:37:39.93 & -38:20:17.2 \\
IRAS 16552$-$3050 & 16:58:27.30 & -30:55:08.1 \\
IRAS 17291$-$2147 & 17:32:11.17 & -21:50:02.4 \\
IRAS 18043$-$2116 & 18:07:20.85 & -21:16:12.1 \\
IRAS 18113$-$2503 & 18:14:26.70 & -25:02:55.8 \\
IRAS 18139$-$1816 & 18:16:49.21 & -18:15:01.7\\
IRAS 18286$-$0959 & 18:31:22.94 & -9:57:21.5 \\
IRAS 18450$-$0148 & 18:47:41.16 & -1:45:11.5\\
IRAS 18455+0448   & 18:48:02.30 & 4:51:30.4 \\
IRAS 18460$-$0151 & 18:48:43.03 & -1:48:30.4 \\
IRAS 18596+0315   & 19:02:06.28 & 3:20:15.7 \\
IRAS 19134+2131   & 19:15:35.21 & 21:36:33.8 \\
IRAS 19190+1102   & 19:21:25.14 & 21:36:33.8 \\
		\end{tabular}
% 	\end{threeparttable}
\end{center}
\end{table*}
% \afterpage{\clearpage}

\begin{table}[!pht]
	\caption{Estimated event rates in the Milky Way.}             % title of Table
	\label{Tab:3}      % is used to refer this table in the text
	\begin{center}
		\begin{tabular}{ c c c}
			\hline\hline
			Mass range & CE events/yr & End of AGB phase/yr    \\
			($M_{\odot})$& &  \\
			\hline 
		0.8-1.2 &0.061 $\pm$ 0.020&0.406 $\pm$ 0.134\\ 
			1.2-2.0 &0.063 $\pm$ 0.021&0.243 $\pm$ 0.080\\ 
			2.0-5.0 &0.017 $\pm$ 0.006&0.047 $\pm$ 0.016\\
			5.0-8.0	&0.007 $\pm$ 0.002	&0.011 $\pm$ 0.004\\
			 Total (0.8-8.0)	&0.148 $\pm$ 0.049&0.706 $\pm$ 0.233\\
			\hline 
		\end{tabular}
	\end{center}
\end{table}
% \afterpage{\clearpage}

\end{document}